\documentstyle[12pt]{article}
\begin{document}
\vskip 2 cm
\begin{center}
\Large{\bf ON STANDARD MODEL HIGGS AND SUPERSTRING THEORIES} 
\end{center}
\vskip 2 cm 
\begin{center}
{\bf AFSAR ABBAS}
\vskip 3 mm
Institute of Physics, Bhubaneswar-751005, India

(e-mail : afsar@iopb.res.in)
\end{center}
\vskip 20 mm
\begin{centerline}
{\bf Abstract }
\end{centerline}
\vskip 3 mm

It is shown that in the Standard Model, the property of charge
quantization holds for a Higgs with arbitrary isospin and hypercharge.
These defining quantum numbers of the Higgs remain unconstrained while the
whole basic and fundamental structure of the Standard Model remains
intact. Hence it is shown that the Higgs cannot be a physical particle.
Higgs is the underlying `vacuum' over which the whole edifice of the
Standard Model stands. Also on most general grounds it is
established here that as per the Standard Model there is no electric
charge above the electro-weak phase transition temperature. Hence there
was no electric charge present in the Early Universe. The Superstring
Theories are flawed in as much as they are incompatible with this
requirement.

\newpage

\section{Charge Quantization in the Standard Model}

 The SM assumes a repetitive structure for each generation of quarks and
leptons \cite{babu}.
Let us start by looking at the first generation \cite{he}
of quarks and leptons
(u, d, e,\,$\nu$ )  and assign them to
$SU(N_{C}) \otimes SU(2)_L \otimes U(1)_Y$ (where $ N_{C} 
= 3 $ ) representation.
To keep things as general as possible this brings in five unknown
hypercharges \cite{a1}. Let us now define the electric charge in
terms of the diagonal generators of $SU(2)_L \otimes U(1)_Y$ as

\begin{equation} Q = T_3 + bY \end{equation}

In the SM $ SU(N_{C}) $ $\otimes$ $ SU(2)_{L}$ $\otimes$
$U(1)_{Y}$ is spontaneously broken through the Higgs mechanism to the
group $ SU(3)_{c} $ $\otimes$ $U(1)_{em}$ . In SM the Higgs is
assumed to be a doublet \cite{a2}. However we do not use this restriction
either and assume the Higgs $ \phi $ to have any isospin T and arbitrary
hypercharge $ Y_{\phi} $. The isospin $ T_{3}^{\phi} $ component of the
Higgs develops a nonzero vacuum expectation value $<\phi>_o$. Since we
want the $U(1)_{em}$ generator Q to be unbroken we require $Q<\phi>_o=0$.
This right away fixes b in (3) and we get

\begin{equation} 
Q = T_3 - ( \frac{ T_{3}^{\phi} }{ Y_{\phi} } ) Y
\end{equation}

For the SM to be renormalizable we require that the triangular anomaly
be canceled. This leads to certain constraints which place some
restrictions on the hypercharges.
Also the use of the fact that after the spontaneous breaking of 
$ SU( N_{ C } ) \otimes SU( 2 )_{L} \otimes U( 1 )_{ Y } $ to 
$ SU( N_{C} ) \otimes U(1)_{ em } $, the L- and R-handed charges couple
identically with photon helps in pinning down hypercharges in terms
of the hypercharge of the Higgs. Hence one obtains quantized
electric charges as 

\begin{displaymath} 
\newline Q(u) = {1\over 2}(1+{1\over N_c})
\end{displaymath}
\begin{equation}  
\newline Q(d) = {1\over 2}(-1+{1\over N_c})
\end{equation}

\begin{eqnarray}
Q(e) = - 1 \\
Q( \nu ) = 0
\end{eqnarray}

For $ N_{C} = 3 $ these are the correct charges in the SM. Note that this
charge quantization in the SM holds for Higgs for arbitrary T and
arbitrary hypercharge. Hence as far as charge quantization is concerned,
the values of T and $ Y_{ \phi } $ remain unconstrained. This point, for
the special case of the Higgs doublet 
was already noted by the author earlier \cite{a2}.

\section{Higgs particle - a ghost !}

 Let us continue with the rest of the structure of the SM and see how our
general Higgs with unconstrained and unspecified isospin T and
hypercharge $ Y_{ \phi } $ fits into it. We can write the covariant
derivative of the SM as

\begin{equation}
D_{\mu} = \partial_{\mu} + i g_{1} \frac{ T_{3}^{\phi} }{ Y_{\phi} }
Y B_{\mu } - i g_{2} \vec{T} . \vec{ W_{\mu} }
\end{equation}

The photon field $ A_{
\mu } $ and the orthogonal $ Z_{ \mu } $ are written as 

\begin{equation}
A_{ \mu } = \frac{ g_{ 2 }B_{ \mu } + g_{1} 
                 ( \frac{ 2 T_{3}^{ \phi } }{ Y_{ \phi } } Y_{l} )
                 W_{ \mu}^{0} }
             { \sqrt{ g_{ 2 }^{2} + 
            ( g_{1} \frac{ 2 T_{3}^{\phi} }{ Y_{\phi} }  Y_{l} )^2 } }    
\end{equation}

\begin{equation}
Z_{ \mu } = \frac{ -g_{1} 
         ( \frac{ 2 T_{3}^{ \phi } }{ Y_{\phi} } Y_{L} ) B_{ \mu } 
             + g_{2} W_{ \mu }^{ 0 } }
             { \sqrt{ g_{2}^{2} + 
             (g_{1} \frac{ 2 T_{3}^{ \phi } }{ Y_{\phi} } Y_{l} )^{2} } }
\end{equation}

With $ D_{ \mu } $ we can write the lepton part of the SM
Lagrangian as 

\begin{eqnarray}
{ \cal{L} }(lepton) = \bar{ q_{L} } i \gamma^{ \mu } 
      ( i g_{1} \frac{ T_{3}^{\phi} }{ Y_{ \phi } } Y_{ l } B_{ \mu } )
       q_{L} 
       + \bar{ e_{R} } i \gamma^{ \mu } 
      ( i g_{1} \frac{T_{3}^{ \phi }}{ Y_{ \phi } } Y_{e} B_{ \mu } )
        e_{R} \nonumber \\
        - \bar{ q_{L} } i \gamma^{ \mu } 
          \left[ i g_{2} \vec{T} . \vec{ W_{\mu} } \right] q_{L}   
\end{eqnarray}

We find that with the hypercharges as specified above the complete
structure of the Standard Model stands intact \cite{a3}.
The point to be emphasized is that all this is independent of Higgs
isospin and hypercharge, which all throughout remain
unconstrained and undetermined. One should not fix any arbitrary values
for them as nothing in the theory demands it.

We find that from rho parameter also \cite{a3}
the solutions for the isospin of the Higgs are infinite in number. 
Again, nothing in the theory demands that one
fixes the isospin to a particular value.

 The point is that the full structure of the SM stands intact without
constraining the quantum numbers isospin and/or the hypercharge of the
Higgs to any specific value. All the
particles that have been isolated in the laboratory or have been studied
by any other means, besides having a specific mass, have definite
quantum numbers which identify them.
In the case of Higgs here, no one knows
of its mass and more importantly its quantum numbers like isospin and
hypercharge, as shown above, are not specified. The
hypercharge of all the other particles are specified as being proportional
to the Higgs hypercharge  which itself remains unconstrained. 
That is, all the hypercharges of particles are rooted on to the Higgs
hypercharge which itself remains free and unspecified. Hence Higgs is
very different from all known particles.
Because of the above reasons Higgs cannot be a physical particle which may
be isolated and studied. It must be just the `vacuum' which sets up the
structure of the whole thing. So Higgs is a ghost which shall not manifest
itself as a genuine particle in the laboratory \cite{a3}. 

\section{Superstring Theories-intrinsically flawed!}

Now  we ask the question, with this generalized
picture what happens to the electric charge when the full Standard 
Model symmetry is restored.

Note that the expression for Q in (2) arose due to spontaneous symmetry
breaking of $ SU(N_{C}) \otimes SU(2)_{L} \times U(1)_{Y} $
(for $ N_{C} = 3 $ ) to 
$ SU(N_{C}) \times U(1)_{em} $ through the medium of a Higgs with
arbitrary isospin T and hypercharge $ Y_{\phi} $. What happens when at
higher temperature, as for example found in the early universe, the 
$ SU(N_{C}) \otimes SU(2)_{L} \otimes U(1)_{Y} $ symmetry is restored ?
Then the parameter `b' in the electric charge definition remains
undetermined. Note that `b' was fixed above due to spontaneous
symmetry breaking through Higgs. Without it `b' remains unknown. 
Hence when the electroweak symmetry is restored, irrespective of the Higgs
isospin and hypercharge the electric charge
disappears as a physical quantity. Therefore we find that there was no
electric charge in the early universe.

Here attention is drawn to the fact
that all putative extensions of the Standard Model should reduce smoothly
and consistently to the Standard Model at low energies. Not only that, all
these extensions should be consistent with the predictions of the Standard
Model at very high temperatures. Contrary to naive expectations, the SM
does make specific predictions at very high temperatures too.
For example one clear-cut prediction of the Standard Model as shown here
and also shown earlier, is that
at high enough temperatures (as in the early universe) when the unbroken 
$ SU(3) \otimes SU(2) \otimes U(1) $ symmetry was restored, there was no
electric charge. GUTs and other standard extensions of the SM are
incompatible with this requirement \cite{a4}.

What about Superstring Theories ? Quite clearly, generically in
Superstring Theories electric charge exists right up to the Planck 
Scale \cite{we}.
Hence as per this theory the electric charge, as an inherent property of
matter, has existed right from the beginning \cite{sch}.
This is not correct in the SM. As shown here and earlier,
the electric charge came into existence at a later stage in the evolution
of the Universe when the  $ SU(2)_{L} \otimes U(1)_{Y} $
group was spontaneously broken to $ U(1)_{em} $. 
It was never there all the time. This is because electric charge is a
derived quantity. Hence we find that in this regard the Superstring
Theories are inconsistent with the SM and hence flawed \cite{a5}.

\end{document}